# Magnetic phase transition induced ferroelectric polarization in BaFeF$_4$ with room temperature weak ferromagnetism

Fan Zhang[a], Yongsen Tang[b], Ranran Li[a], Tianyu Liu[b], Dingshi Xu[a], Yinzhu Chen[a], Ben Niu[b], Shijun Yuan[a], Sai Qin[c], Zhibo Yan[b], Jun Du[b,*], Di Wu[b], Qi Li[a,b], Shuai Dong[a,*], Qingyu Xu[a,b,*]

[a] School of Physics, Southeast University, Nanjing 211189, China

[b] National Laboratory of Solid State Microstructures, Nanjing University, Nanjing 210093, China

[c] School of Sciences, Changzhou Institute of Technology, Changzhou 213032, China

*Corresponding authors: jdu@nju.edu.cn (J. D.), sdong@seu.edu.cn (S. D.),

xuqingyu@seu.edu.cn (Q. X.)

## ABSTRACT

Ba$M$F$_4$ ($M$=Fe, Co, Ni and Mn) family are typical multiferroic materials, having antiferromagnetism at around liquid nitrogen temperature. In this work, polycrystalline BaFeF$_4$ has been prepared by solid state reaction. The slight deficiency of Fe leads to the coexistence of valence states of +2 and +3, facilitating the electrons to hop between the neighboring Fe$^{2+}$ and Fe$^{3+}$ ions through the middle F$^-$ ion, leading to the strong double exchange interaction with weak ferromagnetism above room temperature. A bifurcation at about 170 K between the zero-field-cooled and field-cooled temperature dependent magnetization curves indicates the onset of 2-dimensional antiferromagnetism, which is completed at about 125 K with the sudden drop of magnetization. Despite the fact of type-I multiferroic, its magnetoelectricity can be evidenced by the pyroelectric current, which shows a peak starting at about 170 K and finishing at about 125 K. The saturated ferroelectric polarization change of around 34 μC/m$^2$ is observed, which is switchable by the reversed poling electric field and decreases to about 30 μC/m$^2$ under a magnetic field of 90 kOe. This magnetoelectricity can be qualitatively reproduced by first-principles calculations. Our results represent substantial progress to search for high-temperature multiferroics in ferroelectric fluorides.

# I. INTRODUCTION

Multiferroic materials, possessing ferroelectric and magnetic orderings simultaneously, provide fascinating functions of mutual control of ferroelectricity and magnetism by electric and magnetic fields through the magnetoelectric coupling (ME) [1–3]. However, single-phase multiferroic materials are very rare, since most of the studies are focused on oxide perovskite ferroelectric materials containing 3d transition metal (TM) ions [4–7]. Due to covalent bonding between O and 3d TM ions, the requirements for the outer shell electronic structure for 3d TM ions are contradictory for ferroelectricity and ferro(ferri, or antiferro)magnetism [8]. To break this limit, one possible solution is to search single phase multiferroic materials with more ionic bonding containing 3d TM ions [9].

Ba$M$F$_4$ ($M$=Fe, Co, Ni, and Mn) family are typical ferroelectric materials, which was first reported by Eibschutz *et al*. in the 1960s [10,11]. The ferroelectricity with spontaneous polarization of several μC/cm$^2$ has been reported by the pyroelectric measurements, and the Curie temperature was determined to be much higher than room temperature [11–13]. However, the superexchange interaction between the neighboring 3d TM ions through the middle F$^-$ ions generally leads to the antiferromagnetic coupling. Furthermore, due to the ionic bonding between 3d TM and F$^-$ ions, the overlapping of the electron clouds is rather small, leading to the much weak exchange interaction. Thus, the magnetic structure of Ba$M$F$_4$ is essentially antiferromagnetic, with Néel temperature $T_N$ much lower than room temperature (20-80 K) [14].

The ME in Ba$M$F$_4$ has been preliminarily studied. Weak ferromagnetism has been observed in BaMnF$_4$, which has been attributed to the ferroelectrically induced spin canting between the neighboring Mn$^{2+}$ ions through ME effect [15]. The magnetization in BaMnF$_4$ has dominant orientation after field cooling due to the coupling with spontaneous polarization through ME, and exchange bias effect has been observed [16]. C. Ederer *et al*. has calculated the magnetic structure of BaNiF$_4$, and spin canting between the neighboring Ni$^{2+}$ ions have been reported, which leads to the net magnetic moment. However, due to the symmetry forbidden, the net magnetic moments were aligned antiparallel to each other to form the weak antiferromagnetism. They predicted the electric-field-switchable magnetic order parameter due to its orientation depending on the spontaneous polarization [14]. The weak antiferromagnetism in BaNiF$_4$ has been demonstrated by the wasp-waisted field dependent magnetization hysteresis loops [17]. However, till now, the direct observation of ME effect has not yet been reported.

Motivated by these discussions, it is necessary to realize ferromagnetism above room temperature in Ba*M*F$_4$ materials, and the dependence of ME effect on electric polarization should be directly measured. In this work, BaFeF$_4$ has been investigated since Fe is a typical 3d TM element which always forms ferromagnetic compound with high Curie temperature. By slight deviation from the stoichiometric BaFeF$_4$ with less Fe concentration, a portion of the Fe ions change to be in +3 valence state. Thus, double exchange interaction between the neighboring $Fe^{2+}$ and $Fe^{3+}$ ions can be formed, leading to the ferromagnetic exchange interaction and weak ferromagnetism above room temperature. Furthermore, the electric field reversable polarization change and magnetic field tunable polarization change are observed, which are ascribed to the phase transition to 2-dimensional antiferromagnetism.

## II. EXPERIMENTAL DETAILS

BaFeF$_4$ is synthesized by solid state reaction. The molar ratio of BaF$_2$ and FeF$_2$ powders is finely adjusted to be 1:0.95 to obtain the pure phase sample. The raw materials are mixed and ground for 1 hour in a mortar. The mixture is pressed into small discs in the diameter of 10 mm, which are then put into a copper tube. The tubes are sealed in an arc-melting furnace filled with pure Ar atmosphere (99.99%). Finally, the copper tubes are placed in a furnace and heated at the temperature of 650 ºC for 36 hours. After that, the furnace is naturally cooled down to room temperature, and the final products are obtained.

The crystal structure is studied by X-ray diffraction (XRD, Rigaku Smartlab 3) with Cu Kα radiation ($\lambda$=1.5406 Å). The surface morphology is characterized by a scanning electron microscope (SEM, FEI Inspect F50), and the elemental composition is analyzed by the attached energy dispersive X-ray spectroscope (EDS). The valence state of each element is analyzed by X-ray photoelectron spectroscopy (XPS, PREVAC) with an Al Kα X ray source ($h\nu$=1486.6 eV). Second harmonic generation (SHG) measurements are carried out on a home-built system. The wavelength of excitation laser is 800 nm, which is frequency doubled by a BBO crystal from an fs-laser (Chameleon Compact OPO, center wavelength: 900 nm, repetition rate: 80

MHz), and the integration time is 1s. The excitation laser is focused by a microscopy objective (50×, NA: 0.5). The ferroelectric properties are tested by a scanning probe microscope (SPM, Asylum Research Cypher). The magnetic properties are studied by a superconducting quantum interference device (SQUID, Quantum Design). The $T$ dependence of pyroelectric current ($I_{py}$) is collected upon $T$ increasing from 5 to 300 K using a Keithley 6514 programmable electrometer, with a sample warming rate of 4 K/min. In addition, to explore the ME coupling, the measurement is performed in a physical property measurement system (PPMS, Quantum Design) under a magnetic field of 90 kOe. The variation in electric polarization ($\Delta P$) is obtained by integrating the pyroelectric current with the time. To confirm the electric field reversable $\Delta P$, an electric field of ± 2 kV/cm is applied during the cooling process to 5 K, which is removed during the warming process for the pyroelectric current measurements.

Our density functional theory (DFT) calculations are performed using Vienna *ab initio* Simulation Package (VASP) [18]. The electronic interactions are described by projector-augmented-wave (PAW) pseudopotentials with semicore states treated as valence states [19]. To precisely describe the crystal structure, the generalized gradient approximation (GGA) with Perdew-Burke-Ernzerhof functional modified for solids (PBEsol) parametrization is adopted [20]. To count the correlation effect, the Hubbard $U_{eff}=U-J=4$ eV [21], is applied to Fe's 3d orbitals using the Dudarev approach [22]. Plane-wave cutoff energy is fixed as 500 eV. The k-point grids of 5×3×6 is adopted for both structural relaxation and static computation. The convergent criterion for the energy was set to $10^{-6}$ eV, and that of the Hellman-Feynman forces during the structural relaxation is 0.001 eV/Å. The polarization is calculated using the standard Berry phase method [23,24].

## III. RESULTS AND DISCUSSION

### A. Crystal structure, micromorphology, and elemental analysis

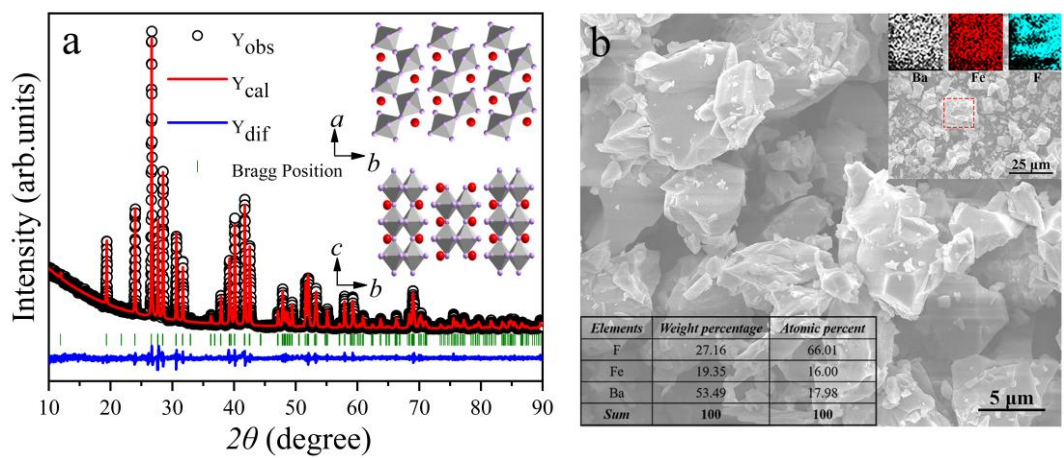

Fig. 1. (a) Rietveld refined XRD pattern of the BaFeF$_4$ powders, the inset shows the schematic diagram of the crystal structure viewed along [001] and [100] orientations. (b) The SEM image of BaFeF$_4$ powders, the insets show the elemental (Ba, Fe, and F) mapping (top right) from the selected area marked by the red dotted box, and the EDS results of the concentration of each element.

It has been reported that the crystal structure of Ba$M$F$_4$ is orthorhombic with [$M$F$_6$] octahedral chain, and the space group is $A2_1am$ [25,26]. The crystal structure of BaFeF$_4$ powders is studied by XRD, and the pattern is shown in Fig. 1a. The Rietveld refinement is applied to get the crystal lattice constants by GSAS-II using the crystallographic data of BaFeF$_4$ (CIF ID#1539611) [27]. The discrepancy factors of refinement are $W_{Rp}$=9.31% and $\chi^2$=3.119. The pure phase is confirmed by that all the peaks can be indexed to the standard data of BaFeF$_4$. The refined results show the orthorhombic structure for the cell with parameters of $a$=5.7680(6) Å, $b$=14.9640(8) Å, and $c$=4.2592(7) Å, and the cell volume is $V$=366.634 Å$^3$. The schematic crystal structure is shown in the inset of Fig.1a, in that the [FeF$_6$] octahedra form the linear chain along the $c$-axis and the puckered sheet along the $a$-axis, and Ba cations are separated by these octahedrons, which agrees well with previous reported data [21,28].

The SEM image of BaFeF$_4$ powders (Fig. 1b) shows the irregular shape with different size. The elemental mappings for Ba, Fe, and F are shown in the inset, and the uniform color indicates the homogeneous distribution. According to the EDS results, the atomic ratio of Fe and Ba is about 0.89, indicating slight deficiency of Fe in BaFeF$_4$ powders, which is due to slightly less amount of FeF$_2$ compared to the stoichiometric composition. Thus, to balance the valence state of each element, valence state of Fe should deviate from +2, and partly in +3, which will be confirmed by the following XPS results.

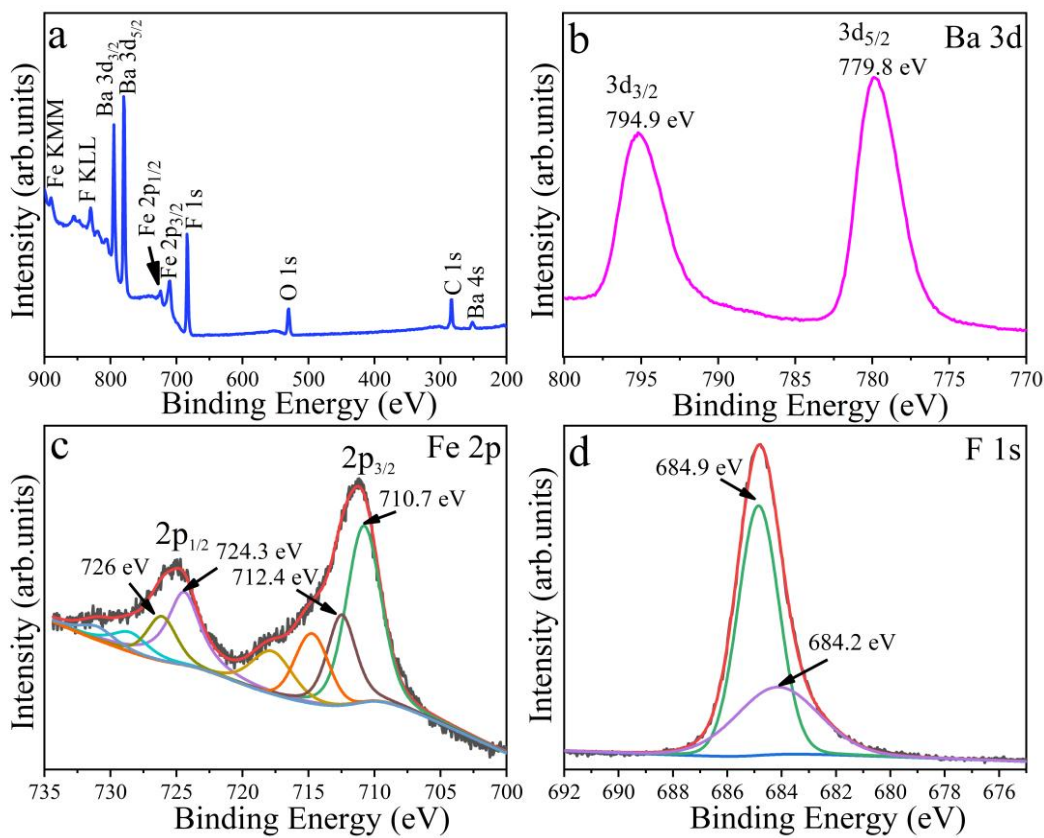

Fig. 2. (a) The full-scan spectra, (b) Ba 3d, (c) Fe 2p, and (d) F 1s XPS spectra of BaFeF$_4$ powders.

XPS measurement is performed to study the valence state of each element, and the collected spectra are shown in Fig. 2. As can be seen, the survey spectrum (Fig. 2a) shows the presence of Ba, C, O, F, and Fe elements in the BaFeF$_4$ powder. Additionally, there are Auger signals assigned to the Fe KMM and F KLL transitions. The C and O may come from the adsorbed impurities in the air, and the C 1s at 284.6 eV is used to correct the instrument error. The Ba $3d_{3/2}$ and $3d_{5/2}$ are at 794.9 eV and 779.8 eV, respectively, leading to the separation of 15.1 eV, which can be attributed to Ba$^{2+}$ ions [29,30]. The high-resolution Fe 2p spectrum is shown in Fig. 2c. The spectrum is deconvoluted into the main peaks from Fe$^{2+}$ and Fe$^{3+}$, together with the corresponding satellite peaks. As can be seen, the fitted curve agrees with the experimental curve quite well. The peaks at 710.7 eV and 724.3 eV are from Fe$^{2+}$, similar to those of FeF$_2$ [31]. The peaks at 712.4 eV and 726 eV confirm the existence of Fe$^{3+}$ ions, due to the nonstoichiometric raw materials with less FeF$_2$ [32]. However, the concentration of Fe$^{3+}$ should be low, due to the much smaller peak area. A rough estimation of Fe$^{3+}$ concentration can be calculated to be about 33% from the area ratio with $Fe^{2+}$ of 1:2, if we assume the equal scattering factor of $Fe^{2+}$ and $Fe^{3+}$ ions. This is quite consistent with the EDS result of 11% Fe deficiency, which will make 25% Fe in +3 valence state if the F concentration keeps constant. The XPS peak of F 1s can be deconvoluted to two peaks. The peak at 684.9 eV can be ascribed to the Fe-F bonding in stoichiometric area [33]. And, the other peak at 684.2 eV can be attributed to Fe-F bonding in Fe deficient area. The area ratio of the F 1s peak at 684.2 eV to that at 684.9 eV is also 1:2, which is consistent with the area ratio of $Fe^{3+}$ to $Fe^{2+}$.

## B. Nonlinear optical behavior, magnetic phase transitions, and ferroelectric polarization

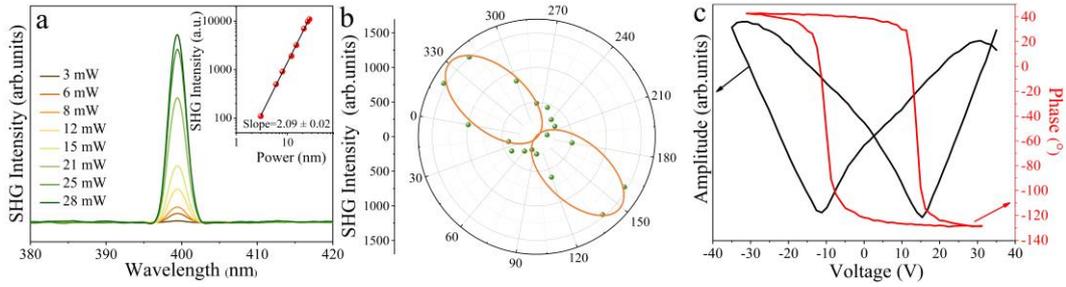

Fig. 3. SHG results are measured at the room temperature. (a) SHG intensity of $BaFeF_4$ with different excitation power, the inset is the excitation power-dependent SHG intensity in logarithmic coordinates, (b) Polarization-angle-dependence of the SHG intensity, (c) Phase–voltage and amplitude–voltage piezoelectric curves at room temperature.

The spontaneous polarization of $BaFeF_4$ has been confirmed by the pyroelectric current measurement, but the reverse of polarization by electric field has not yet been reported, which has been ascribed to the much larger radius of $Fe^{2+}$ ions, leading to the too high energy barrier [21]. $BaFeF_4$ has a polar space group, which can be demonstrated by the SHG measurement. The excitation wavelength is 800 nm, and the corresponding SHG signal appears at 400 nm. The SHG intensity increases with the enhancement of power, as shown in Fig. 3a. The inset of Fig. 3a shows the SHG intensity in dependence on the power intensity both in Log scale, and the linear fitting shows the slope of about 2.09, which is very close to the theoretical value of 2 [34]. This indicates that SHG intensity is proportional to the square of the power intensity, demonstrating nonlinear optical principal [35,36]. Since the polarization-resolved SHG shows crystal-symmetry dependence, we perform the angle dependent SHG intensity with excitation laser parallel to polarization. The polarization-dependent SHG intensity in Fig. 3b shows a twofold rotational symmetry by varying the azimuthal angle $\theta$, with its minimum at 60° and maximum at 150°, which also exhibits the nonlinear behavior.

We perform the electric field dependent polarization $P(E)$ measurement to check the ferroelectric properties of $BaFeF_4$. However, the large leakage current impedes the observation of well-shaped $P(E)$ loops. To avoid the severe leakage problem, we use piezoresponse force microscopy (PFM) by SPM taken in the normal mode of atomic force microscopy (AFM) to confirm the ferroelectric properties. Clear butterfly shaped phase–voltage and amplitude–voltage hysteresis loops can be observed, as shown in Fig. 3c, indicating the ferroelectric properties of $BaFeF_4$. The nearly 180º phase change indicates the total reverse of ferroelectric polarization might be realized, due to that much higher electric field can be applied from the acuate tip with very small radius.

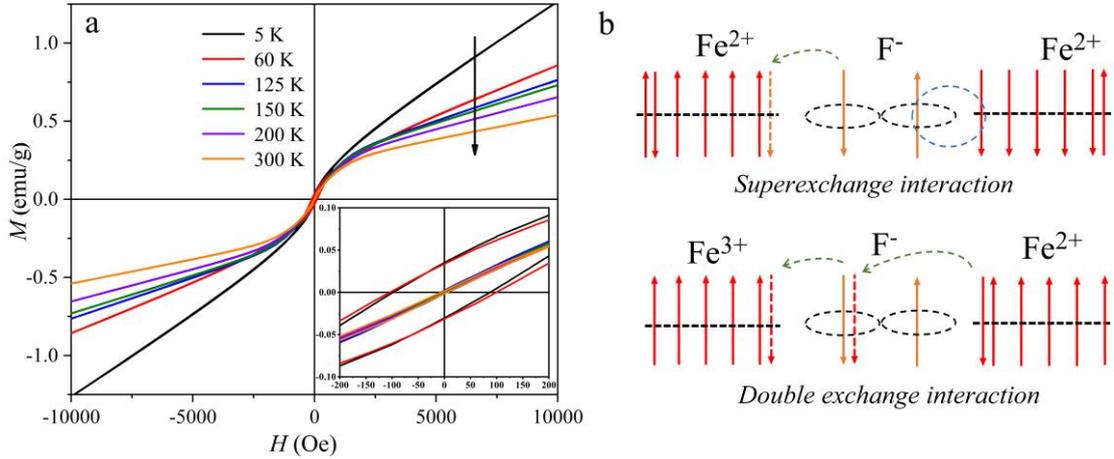

Fig. 4. (a) Magnetic field dependence of magnetization $M(H)$ at different temperature cooled under the magnetic field of 1000 Oe, the inset is the enlarged view. (b) The schematic diagram for the superexchange interaction between $Fe^{2+}$ ions and the double exchange interaction between $Fe^{2+}$ and $Fe^{3+}$ ions.

The macroscopic magnetism is characterized by measuring the $M(H)$ curves at various temperatures, as shown in Fig. 4a. It can be clearly seen that S-shaped $M(H)$ curves can be observed up to 300 K, in contrast to the previous reported linear $M(H)$ curves for $BaMnF_4$, $BaNiF_4$, and $BaCoF_4$ [16,17,37]. However, a small coercivity of 7 Oe is observed for the $M(H)$ curve at 300 K, indicating the soft ferromagnetism in $BaFeF_4$. With further decreasing temperature to below 60 K, the $M(H)$ curves show clearly the hysteresis loop with significantly large coercivity. To further understand the magnetic properties of BaFeF$_4$, zero field cooled (ZFC) and field cooled (FC) temperature dependent magnetization $M(T)$ curves are measured, as shown in Fig. 5a. For FC process, a magnetic field of 200 Oe is applied. The FC magnetization increases slightly with decreasing temperature to around 125 K. Then a drastic decrease of magnetization can be observed. With further decreasing temperature, a hump can be observed with peak position at about 56 K. Below about 20 K, the magnetization increases drastically with decreasing the temperature. The ZFC magnetization shows similar trend with FC magnetization, but a bifurcation can be observed at about 170 K between the FC and ZFC $M(T)$ curves. With further decreasing temperature, ZFC magnetization shows the drastic drop at slightly higher temperature than FC magnetization. Ba$M$F$_4$ has layered structure, with puckered sheet in $ac$ plane. Due to the larger separation between the neighboring layers, the interlayer exchange interaction is weaker than the intralayer exchange interaction. Thus, 2-dimensional antiferromagnetism is formed inside each layer first with decreasing temperature (e.g., BaMnF$_4$ (50 K), BaNiF$_4$ (150 K), BaCoF$_4$ (95 K) [16,17,37]). With further decreasing temperature, 3-dimensional antiferromagnetism is formed. Though the 2-dimensional antiferromagnetic transition in BaFeF$_4$ has not been reported, it is reasonable to attribute the sudden drop of magnetization at about 125 K to the form of 2-dimensional magnetic structure. The hump at 56 K, close to the previously reported $T_N$ of about 60 K for BaFeF$_4$, can be attributed to the 3-dimensional antiferromagnetic transition [10]. The separation between FC and ZFC $M(T)$ curves starts at 170 K, indicating that the intralayer antiferromagnetic exchange interaction emerges at 170 K. But in the beginning, only antiferromagnetic clusters formed in the paramagnetic bulk, due to the slight inhomogeneous distribution of each element, or electronic fluctuation. With further decreasing temperature, the antiferromagnetic clusters expand and finally connect together, then 2-dimensional antiferromagnetic structure is completely formed.

Interestingly, the magnetization only shows gradual decrease with increasing temperature up to 300 K, and together with the S-shaped $M(H)$ curve at 300 K with coercivity of 7 Oe, indicating the existence of room temperature weak ferromagnetism. In our previous work on Sr$_3$Fe$_2$F$_{12}$, Ba$_5$Fe$_3$F$_{19-\delta}$ and Pb$_5$Fe$_3$F$_{19}$ prepared using the same raw materials and under the similar conditions, only linear $M(H)$ curves have been observed at 300 K, excluding the possible ferromagnetic impurities under the detecting limit [29,38,39]. This is abnormal, since such ferromagnetism has not been observed at 300 K in the bulk $BaMnF_4$, $BaCoF_4$, and $BaNiF_4$ [16,17,37]. In stoichiometric $BaFeF_4$, Fe should be in +2 valence state. As can be seen in Fig. 4b, due to the ionic bonding between $Fe^{2+}$ and $F^-$ ions, the overlapping of electron clouds is quite small, leading to the much weak exchange interaction and low magnetic ordering temperature. Furthermore, only superexchange interaction between the $Fe^{2+}$ ions can be established, leading the antiferromagnetism in $BaFeF_4$. As confirmed by the XPS results of the existence of $Fe^{3+}$ ions, double exchange interaction might be formed between the neighboring $Fe^{2+}$ and $Fe^{3+}$ ions. The outer shell electronic structure of $Fe^{2+}$ is $3d^6$, while that of $Fe^{3+}$ is $3d^5$. An electron in $F^-$ can jump to the neighboring $Fe^{3+}$ with spin down antiparallel to the local d spins due to the half-filled nature of $Fe^{3+}$. Then, an electron with spin down in $Fe^{2+}$ can jump to $F^-$ to occupy the vacancy. This is called double exchange interaction. After this process, the $Fe^{3+}$ ion becomes $Fe^{2+}$, while $Fe^{2+}$ changes to $Fe^{3+}$. Thus, the electronic structure before and after are the same, indicating the degenerated states. This will not only lead to the ferromagnetic exchange interaction between the neighboring $Fe^{3+}$ and $Fe^{2+}$ ions, but also enhance the interaction strength due to the increased electron hopping possibility. Thus, weak ferromagnetism can be established even up to room temperature. However, due to the nonuniform distribution of $Fe^{3+}$ ions, strong ferromagnetic interaction can only be established at the regions with local high concentration of $Fe^{3+}$ ions. And the low concentration of $Fe^{3+}$ ions between these regions induces the weak exchange interaction, which are easily reversed by the magnetic field, leading to the low coercivity and soft ferromagnetism at high temperature.

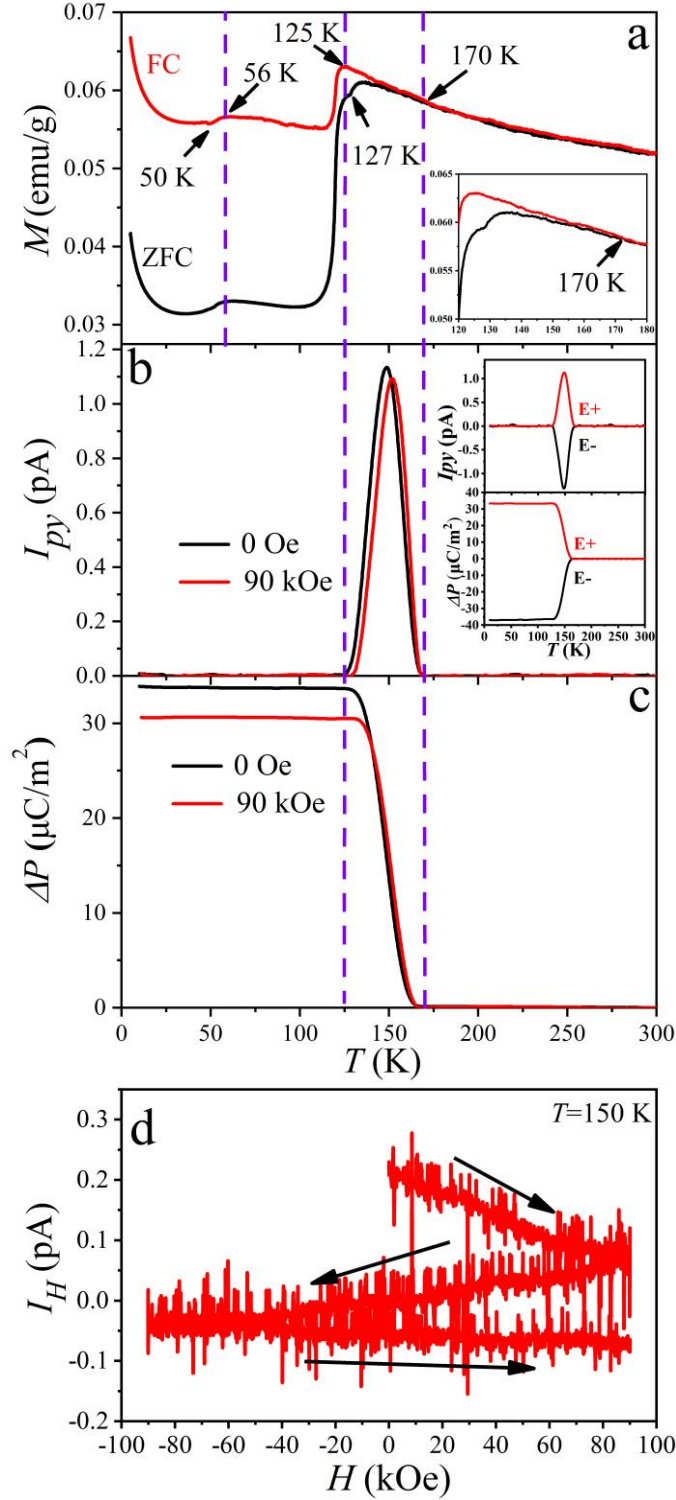

Fig. 5. (a) The FC and ZFC magnetization $M(T)$ curves under a magnetic field of 200 Oe, the inset is the enlarged view. (b) The temperature dependent pyroelectric currents $I_{py}(T)$ curves under different magnetic field. The inset shows the curves of $I_{py}(T)$ and integrated polarization change $\Delta P(T)$ under the positive and negative poling electric field without the magnetic field. (c) The $\Delta P(T)$ curves under different magnetic field. (d) The relationship between the magnetoelectric current $I_H$ and the magnetic field $H$ at $T$=150 K.

The ME effect in the multiferroics is typically manifested as controlling ferroelectric polarization by magnetic field or magnetization by electric field. However, as stated above that the $P(E)$ loop can't be obtained due to the leakage problem, pyroelectric current measurement is applied to get the information of ferroelectric polarization. An electric field of 2 kV/cm is applied during the cooling process to temperature of 5 K, and the current is collected during the warming process with increasing temperature rate of 4 K/min. As BaFeF$_4$ is ferroelectric with Curie temperature of 1093 K [40], a continuous increase of $I_{py}$ can be observed with increasing temperature, which has been subtracted as background signal. As can be seen in Fig. 5b, a sudden increase of $I_{py}$ can be observed at about 170 K, with peak position at about 150 K. $I_{py}$ decreases to around zero at about 125 K. Under temperature below 125 K, no peak of $I_{py}$ can be observed. Thus, besides the change of $I_{py}$ with increasing temperature, the observation of peak for $I_{py}$ indicates an extra polarization change in the temperature range between 125 K and 170 K. The polarization change $\Delta P$ is calculated by integrating the $I_{py}$ with the time, as shown in Fig. 5c. $\Delta P$ starts to emerge at about 170 K, and saturate at about 125 K to ~34 μC/m$^2$. As studied for the magnetic properties, the onset of $\Delta P$ at 170 K coincides well with the onset temperature of 2-dimenional antiferromagnetism. And the finish of the polarization at 125 K is quite close to the finish of the 2-dimensional antiferromagnetic transition. Thus, it is reasonable to attribute $\Delta P$ to the 2-dimensional antiferromagnetic transition. We further repeat the pyroelectric current measurement under a magnetic of 90 kOe. The peak position slightly shifts to higher temperature, and the peak height becomes smaller, indicating the possible ME effect. The calculated $\Delta P$ saturates at smaller value of about 30 μC/m$^2$. We also perform the measurement of the magnetic field dependence of magnetoelectric current $I_H$, but no clear relation can be observed, as shown in Fig. 5d [7]. Similar results are observed at 10 K and 60 K. This might be due to the antiferromagnetic structure, which cannot be effectively manipulated by the applied magnetic field. To break the antiferromagnetic structure, much higher magnetic field is needed. To confirm whether $\Delta P$ can be reversed by an electric field, we further perform the pyroelectric current measurement after the application of electric field of -2 kV/cm during the cooling process. As shown in inset of Fig. 5b, a nearly mirrored $I_{py}(T)$ curve can be observed, and $\Delta P$ is reversed by the reversed poling electric field with saturated $\Delta P$ of -

36 µC/m$^2$.

## C. The DFT calculation

To understand the pyroelectric polarization and magnetoelectricity, a DFT calculation is performed. With the space group *A2$_1$am*, the DFT optimized lattice constants for BaFeF$_4$ is *a*=5.7252 Å, *b*=14.9641 Å, and *c*=4.2622 Å, which is highly consistent with our experimental results and ensures the accuracy of following calculation of ferroelectricity. This space group allows a ferroelectric polarization along the *a*-axis. To precisely understand the experimental Δ*P*, the polarization difference between the antiferromagnetic state and paramagnetic state should be accounted. However, the paramagnetic state cannot be treated in standard DFT calculation. A less-than-ideal alternative is using the ferromagnetic state to replace the paramagnetic state, which can qualitatively mimic the magnetoelectricity, as done in other pyroelectric systems [41].

Our DFT result of polarization change due to antiferromagnetic transition is 1.704 µC/cm$^2$, qualitatively supporting the experimental magnetoelectricity. The quantitative deviation can be understood as following. First, our sample is polycrystalline, with randomly-oriented crystalline grains, which will reduce the measured polarization for at least one order of magnitude. Second, the grain boundaries in polycrystalline samples will reduce the effective poling voltage. Also, the real ferroelectric transition temperature is far above room temperature, thus our poling field starting at 300 K is not capable to generate a single ferroelectric domain state with a saturated polarization. Namely, the experimental Δ*P* should be much smaller than the theoretical saturated value. Therefore, larger Δ*P* is expectable in further studies with single crystals and higher poling fields. Last, the ferromagnetic state used in calculation as the baseline is different from the experimental paramagnetic one.

# IV. CONCLUSIONS

In summary, polycrystalline multiferroic fluoride BaFeF$_4$ has been prepared by solid state reaction. The S-shaped *M*(*H*) curves is observed at 300 K with coercivity of 7 Oe, indicating the existence of room temperature weak ferromagnetism, due to double exchange interaction between the $Fe^{2+}$ and $Fe^{3+}$ ions through the $F^-$ ion. The temperature dependent pyroelectric current shows a peak at about 150 K which starts at 170 K and completes at 125 K. These temperatures coincide well with the onset of 2-dimensional antiferromagnetism at 170 K and finish of this magnetic phase transition at about 125 K. The DFT calculation confirms the polarization change due to the antiferromagnetic phase transition. The ME effect is confirmed by the decrease of saturate ferroelectric polarization change from ~34 $\mu C/m^2$ to ~30 $\mu C/m^2$ after applying a magnetic field of 90 kOe. The ferroelectric polarization change is also switchable by the reversed poling electric field. The observation of ME effect with the related mechanism to the magnetic phase transition sheds light on realizing the room temperature multiferroic fluorides for the practical applications.

# ACKNOWLEDGMENTS

This work is supported by the National Natural Science Foundation of China (51771053, 51971109, 51802031, 11834002), the Fundamental Research Funds for the Central Universities (2242020k30039), and the open research fund of Key Laboratory of MEMS of Ministry of Education, Southeast University.